\documentclass[jcp,onecolumn,preprint,floatfix]{revtex4-1}

\usepackage{graphics} 
\usepackage{graphicx} 
\usepackage{color} 
\usepackage{bm} 
\usepackage[dvipsnames]{xcolor,colortbl}
\usepackage{amsmath}
\usepackage{amssymb}
\usepackage{amsfonts}

\definecolor{DarkYellow}{RGB}{80, 80, 0}

\begin{document} 
\title{Deformable hard particles particles confined in a disordered porous matrix}
\author{Alexander Stadik} 
\author{Gerhard Kahl} 

\affiliation{Institute for Theoretical Physics and Center for
  Computational Materials Science (CMS), Technische Universit\"at Wien, Wiedner Hauptstra{\ss}e 8-10, A-1040 Wien, Austria}


\begin{abstract}
With suitably designed Monte Carlo simulations we have investigated the properties of mobile, impenetrable, yet deformable particles that are immersed into a porous matrix, the latter one realized via a frozen configuration of spherical particles. By virtue of a model put forward by Batista and Miller [Phys. Rev. Lett. {\bf 105}, 088305 (2010)] the fluid particles can change under the impact of their surrounding (i.e., either other fluid particles or the matrix) their shape within the class of ellipsoids of revolution; such a change in shape is related to an energy change which is fed into suitably defined selection rules in the deformation ``moves'' of the Monte Carlo simulations. This concept represents a simple, yet powerful model of realistic, deformable molecules with complex internal structures (such as dendrimers or polymers). For the evaluation of the properties of the system we have used the well-known quenched-annealed protocol (with its characteristic double average prescription) and have analysed the simulation data in terms of static properties (radial distribution function and aspect ratio distribution of the ellipsoids) and dynamic features (notably the mean squared displacement). Our data provide evidence that the degree of deformability of the fluid particles has a distinct impact on the aforementioned properties of the system.
\end{abstract}

\date{\today}

\maketitle


\section{Introduction}
\label{sec:introduction}

Experimental, theoretical, and computer simulation based investigations on fluids confined in disordered, porous confinement have received over meanwhile several decades a steadily increasing share of interest \cite{Gelb:1999,Rosinberg:1999,McKenna:2003,Alcoutlabi:2005}. This is certainly due to the fact that the scenario of confined fluids is ubiquitous in our daily lives, with examples ranging from biophysics over chemical engineering to technological applications. But also from the pure academical point of view such systems have been (and still are) studied extensively in fundamental science which is due to the fact that confined fluids drastically change their properties once they are exposed to the external field of a disordered matrix: this applies not only to the static properties (such as the structure, the thermodynamics, or the phase behaviour) but also to the dynamic properties of the fluid (such as the mean squared displacement of the particles -- and hence the diffusivity --, or the dynamic structure factors).  

From a theoretical point of view -- be it within theoretical frameworks or in computer simulations -- such investigations are particularly challenging both from the conceptual as well as from the numerical point of view: it is difficult to properly define theoretical concepts that allow the reliable evaluation of thermodynamic properties of such systems; and it is also challenging to put forward models -- both for the liquid as well as for the porous confinement -- that faithfully mimic the features of real systems (such as He confined in aerogels  \cite{Wong:1990,Chan:1996}). 

In many of the preceding theoretical investigations the seminal quenched-annealed (QA) model was used: here the confining environment -- henceforward termed matrix (introducing the index 'm') -- is modeled as an instantaneously frozen configuration of an equilibrated liquid; further, the space left void by this matrix is filled by mobile, fluid particles (with index 'f'). This concept represents a viable compromise for the above mentioned requirements: it is amenable both to computer simulations and theoretical frameworks, while it still captures relevant features of realistic systems at a rather faithful level. The related theoretical, statistical mechanics based framework was pioneered by Madden and Glandt \cite{Madden:1988,Madden:1992} and later by Given and Stell \cite{Given:1992,Given:1992a,Given:1994}: within these concepts the system is considered as a very peculiar mixture of the matrix and of the fluid particles. Observable quantities are calculated via a double-averaging procedure: in a first step physical properties are obtained via averaging over the degrees of freedom of the fluid for a given matrix configuration; subsequently the obtained values are averaged over all possible matrix configurations that are compatible with the macroscopic system parameters. Within this formalism the correlation functions between the particles are related via an Ornstein-Zernike type equations (the so-called Replica Ornstein-Zernike equations) which can be solved numerically in combination with a suitable closure relation. This approach provides furthermore access to the thermodynamic properties of the system \cite{Rosinberg:1994,Kierlik:1995,Kierlik:1997,Paschinger:2000,Paschinger:2001,Schmid:2002}. And it should be mentioned that Krakoviack \cite{Krakoviack:2005,Krakoviack:2007,Krakoviack:2009} successfully merged this formalism with the concept of mode coupling theory \cite{Goetze:2009}, paving thus the way to investigate also dynamic properties of confined fluids. Computer simulations, on the other hand, literally execute the above outlined double averaging procedure: typically five to 20 different, but equivalent matrix configurations are selected to realize the ``outer'' averaging procedure. All in all, numerous studies have been performed during the past decades on QA systems to investigate their static \cite{Lomba:1993,Vega:1993,Meroni:1996,Paschinger:2000,Bores:2014,Bores:2015}, thermodynamic \cite{Rosinberg:1994,Kierlik:1995,Kierlik:1997,Alvarez:1999,Paschinger:2001}, or dynamic \cite{Gallo:2003,Kim:2003,Chang:2004,Mittal:2006,Kurzidim:2009,Kurzidim:2010,Kim:2009,Kurzidim:2011} properties. 

In most of these investigations both types of particles are assumed to be spherically symmetric and fixed in their shape. This assumption is presumably not too restrictive for the matrix particles as they are fixed in their positions, anyhow. However the situation is different for the mobile, fluid particles: in particular in the realm of soft matter physics these particles are usually viewed as coarse-grained, ``effective'' models whose shape can in principle vary as a consequence of the internal dynamics of the constituent entities of such molecules. In the vast majority of scientific investigations such ``effective'' particles are considered as spherical; still, in a few contributions the non-spherical shape of these molecules has been taken into account, for instance, in polymers \cite{Murat:1998,Lim:2016,Eurich:2007,Dadamo:2012}, dendrimers \cite{Georgiou:2014,Weissenhofer:2018}, or patchy particles \cite{Bianchi:2015}. Within  the framework of ``effective'' particles it is therefore not surprising that the shape of such ``effective'' particles adopt their shape to their immediate surrounding -- be it another mobile particle or the rigid matrix.

In an effort to take this feature within the QA scenario properly into account it is more appropriate to introduce to a certain degree flexibility in the shape of the mobile particles. In this contribution we address this issue for the first time, taking benefit of the availability of a simple, yet realistic and powerful model of particles of variable shape: our investigations are based a model of deformable, impenetrable particles (proposed by Batista and Miller \cite{Batista:2010,Batista:2011}) which are allowed to change -- at the cost of some energy penalty -- their shape in computer simulations: to be more specific, an initially spherical particle can be deformed into an ellipsoid of revolution, characterized by its aspect ratio $x$. As specified in Refs. \cite{Batista:2010,Batista:2011} the energy penalty for deformation is proportional to  $\beta \kappa  \ln^2 x$, where $\beta \kappa$ is the deformability (or stiffness) of the particles ($\beta$ being the inverse temperature). Batista and Miller have shown that this model is readily amenable to Monte Carlo simulations, introducing suitable, additional rotational and particle deformation ``moves.'' 

We have integrated this model into the framework of QA systems: the mobile, impenetrable particles are now deformable, based on the above outlined model of Batista and Miller.  We apply in our Monte Carlo simulation in an alternating order the following MC ``moves'' to the fluid particles: (i) conventional particle displacement, (ii) rotational moves of the ellipsoidal particles, and (iii) deformation moves. Now that the particles are no longer spherical the verification of the overlap criterium is computationally more cumbersome and considerably more time consuming than in the spherical case: here we have taken benefit of a concept proposed by Vieillard-Baron \cite{Vieillard-Baron:1972} in his seminal work and have implemented the related, numerically very efficient algorithm proposed by Perram and Wertheim \cite{Perram:1985}. For the ``outer'' average of the above mentioned double averaging procedure within the QA framework we have assumed -- as a consequence of the high computational cost of the extended steps -- five independent matrix configurations. Particle ensembles typically range from 3000 to 4000 particles. 

In order to quantify our findings we have focused on the following features: the static radial distribution function and the distribution of the aspect ratio of the fluid particles, and their mean squared displacement. Investigations have been carried out over a significant number of states in the parameter space, spanned by the number density of the matrix particles ($\Phi_{\rm m}$) and of the fluid particles ($\Phi_{\rm f}$) and considering a broad range of deformability parameters $\beta \kappa$. 

The manuscript is organized as follows: in the subsequent Section we present our model and provide details about the simulation technique; we furthermore define four pathways through parameter space along which we have investigated properties of our system. The relevant data are discussed in Section III, with particular emphasis on the radial distribution function of the deformable, mobile particles, the emerging probability distributions of the aspect ratios of the particles, and on their mean squared displacement. The manuscript is closed with concluding remarks and an outlook to future work.

\section{Model and Methods}
\label{sec:model_methods}

\subsection{Model}
\label{subsec:model}

We consider initially an ensemble of $N$ hard spheres of diameter $\sigma$, confined in a volume $V$ at a temperature $T$ (with $\beta = 1/(k_{\rm B} T)$, $k_{\rm B}$ being the Boltzmann constant); $\sigma$ is the unit length of the system and is henceforward set to unity. $N_{\rm m}$ of these particles are used to form the matrix and are henceforward kept fixed in their positions. Furthermore, these matrix particles are undeformable, i.e., they keep their spherical shape; their packing fraction, $\phi_{\rm m}$, is defined as 

$$
\phi_{\rm m} = \frac{\pi}{6} \frac{N_{\rm m}}{V} .
$$ 
The remaining $N_{\rm f} =( N - N_{\rm m})$ fluid particles are immersed into the space left void by the matrix configuration; these particles are mobile and deformable: following the concept of Batista and Miller \cite{Batista:2010,Batista:2011} they can -- subject to an energy penalty -- deform into (still impenetrable) prolate or oblate ellipsoids of revolution. Their packing fraction is defined by 

$$
\phi_{\rm f} = \frac{\pi}{6} \frac{N_{\rm f}}{V} .
$$

The framework behind our model is that of a quenched-annealed (QA) system, which has been designed to investigate the properties of fluid particles immersed into a confinement realized via a disordered particle arrangement: the matrix is assumed to be an instantaneously frozen configuration of an equilibrated fluid with its voids being filled by fluid particles \cite{Given:1992,Given:1994,Rosinberg:1994,Kurzidim:2009,Kurzidim:2011}. The concept of QA systems allows us to calculate thermodynamic averages of observables: working in the canonical ensemble (and following the ideas of previous contributions) the properties of the system (in particular its static and dynamic behaviour) are calculated via a double averaging procedure (see, e.g. \cite{Kurzidim:2009, Kurzidim:2011}): assuming a fixed configuration of matrix particles we average  some property (say $A$) over the degrees of freedom of the mobile fluid particles, leading to the ensemble average $\langle A \rangle$. This procedure is repeated by considering (formally) all different, but equivalent matrix configurations that are compatible with the macroscopic parameters (such as temperature, volume, or number of matrix particles). Averaging over the resulting $\langle A \rangle$-values, leads to the double averaged property $ \bar A = \overline{\langle A \rangle}$. The actual realization of this averaging procedure in Monte Carlo simulations is specified in the subsequent section.

The concept of deformable hard spheres rests on an idea put forward by Batista and Miller \cite{Batista:2010}: starting from spheres with diameter $\sigma$ the particles can change their shape and can become ellipses of revolution (spheroids), with semi-axes $a = b \ne c$; it must be emphasized that during these shape transformations the volume of the particles is preserved, therefore the particles can be considered as incompressible. For convenience we introduce the aspect ratio $x$, defined by $x = a/c$; thus particles are oblate for $x < 1$, prolate for $x > 1$, and spherical for $x = 1$. Consequently particles are not only characterized by their positions ${\bf r}_i$ but also by their orientations ${\mathbf \Omega}_i$ ($i = 1, \dots, N_{\rm f}$) where the orientational unit vectors point along the $c$-axes of the particles. For the different types of Monte Carlo moves (specified below) it is relevant whether two spheroids (located at positions ${\bf r}_1$ and ${\bf r}_2$ and with orientations ${\mathbf \Omega}_1$ and ${\mathbf \Omega}_2$) show after such a move an overlap or not: in the former case, the particle move is rejected while it is otherwise accepted. For spherical particles the implementation of such a criterion is trivial; however, this is not the case for spheroids: in our program we have implemented a reliable and numerically efficient method for identifying the possible overlap of two ellipsoids, put forward by Perram and Wertheim \cite{Perram:1985}. This method is numerically more attractive than an implementation of the original criterium put forward by Vieillard-Baron \cite{Vieillard-Baron:1972}. For the application of this criterium the spherical (matrix) particles are considered as spheroids with $x \equiv 1$. 

\subsection{Monte Carlo simulations}
\label{subsec:MC}

Our canonical Monte Carlo simulations were carried out in a cubic cell, assuming periodic boundary conditions. Three types of ``moves'' were applied to the particles:

\begin{itemize}
    \item[(i)] conventional particle displacements; the maximum value of displacement is adjusted dynamically during the simulation to guarantee an acceptance rate of the translational moves of 50 \%;
    \item [(ii)] rotational moves of the particles; in a trial rotation of a particle its Euler angles are changed by angular increments, whose maximum values are dynamically adjusted to ensure an acceptance rate of the rotational moves of 50 \%;
    \item[(iii)] shape deformation ``moves'' of the particles; following the concept of Batista and Miller \cite{Batista:2010,Batista:2011}, the potential $U(x)$ which governs the particle deformations is given (in lowest order approximation) by 
    
    \begin{equation}
    U(x) = \kappa  (\ln x)^2 ;
    \label{eq:potential}
    \end{equation}
    $x$ is the above introduced aspect ratio of the particles and $\kappa$ is the stiffness (or deformability) parameter; summarizing the details laid out in \cite{Batista:2010}, $\kappa$ is related to the surface tension $\gamma$ of the particles via $\kappa = \frac{8}{45} \sigma^2 \pi  \gamma$. 
    
    In a shape deformation ``move'' the $\ln x$-term in Eq. (\ref{eq:potential}) is changed by an additive ``displacement'' $\Delta x_{\rm shape}$; this value is dynamically adjusted to ensure an acceptance rate of the shape deformation ``move'' of 50 \%; for details we refer to Refs. \cite{Batista:2010,Batista:2011}.
\end{itemize}

Acceptance of all three ``moves'' is governed by the aforementioned overlap criterion: thus, if a trial move leads to an overlap between two particles, this move is definitely rejected. For the shape deformation ``move'' an additional, energy-based criterion is imposed which reads (for details cf. Refs. \cite{Batista:2010,Batista:2011}):

\begin{equation}
p = \min [1, \exp(-\beta U(x))] .
\end{equation}
Thus a shape deformation ``move'' is accepted -- provided that no particle overlap occurs for the deformed particles -- by a probability $p$ specified in the above equation. 

In our notation a sweep consists of a series of subsequent $N_{\rm f}$ translational, $N_{\rm f}$ rotational, and of $N_{\rm f}$ shape deformation ``moves'' of randomly selected fluid particles. At this point it should be emphasized that the check for particle overlap is computationally quite expensive (imposing thus limitations on the ensemble size). 

In view of the above said, our system is characterized by three parameters: the deformability $\kappa$ and the number of fluid ($N_{\rm f}$) and matrix ($N_{\rm m}$) particles. With our choice of $\kappa$-values (i.e. $\beta \kappa =1, 5, 10$ and 100) we cover a broad spectrum ranging from strongly deformable particles (with low $\beta \kappa$-values) to undeformable particles (with $\beta \kappa = 100$); the case of spherical hard spheres is recovered for $\beta \kappa \to \infty$. 

Simulations have been carried out in ensembles of typically 3000 to 4000. The actual values of $N_{\rm f}$ and $N_{\rm f}$ depended on the location of the investigated state in the parameter space spanned by $\Phi_{\rm f}$ and $\Phi_{\rm m}$: in an effort to guarantee a sufficient numerical accuracy of the data we considered in general a number $N_{\alpha} = 4000 \times \Phi_\alpha$ ($\alpha$ standing for 'f' or 'm') of either particle species: thus, for instance, for a system with $\Phi_{\rm m} = 0.20$ and $\Phi_{\rm f} = 0.30$ we considered $N_{\rm m} = 800$ matrix and $N_{\rm f} = 1200$ fluid particles. Only for low-density states (i.e., if either of the densities was 0.05) we used $N_\alpha = 8000 \times \Phi_\alpha$ ($\alpha$ standing for 'f' or 'm'), instead.

\begin{figure}[htbp]
\begin{center}
\includegraphics[width=10cm]{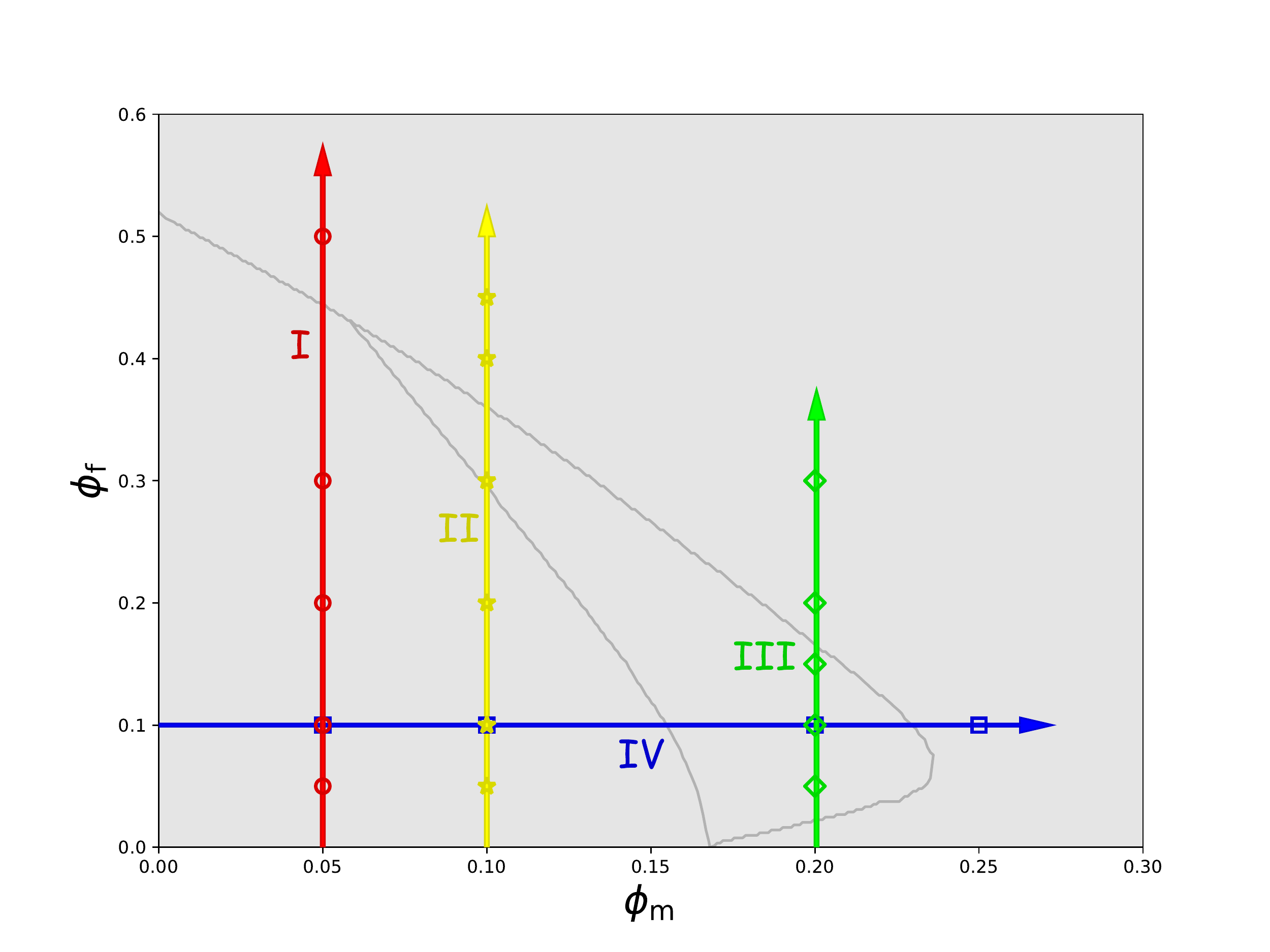}
\caption{(color online) States investigated in the present study (marked by the coloured symbols), shown in the ($\Phi_{\rm m}, \Phi_{\rm f}$)-plane. The four different pathways (labeled I to IV and defined in the text) are shown in different colours. The grey lines sketch the ``dynamic'' phase diagram of a previous study \cite{Kurzidim:2009} where the dynamic properties of spherical particles, confined in a porous matrix, formed by undeformable, spherical particles, were investigated.}
\label{fig:pathways}
\end{center}
\end{figure}

In an effort to scan the relevant regions of the parameter space spanned by $\Phi_{\rm m}$ and $\Phi_{\rm f}$ we have defined four pathways (denoted by I, II, III, and IV); for this choice we were guided by the diagram of states of {\it spherical} fluid particles confined in a matrix of immobile, {\it spherical}, particles, as depicted in Fig. 1 of Ref. \cite{Kurzidim:2009}. These four pathways are shown -- along with symbols representing the specific, investigated states -- in Fig. \ref{fig:pathways} and are defined as follows:

\begin{itemize}
    \item {\bf path I}: the matrix has a small packing fraction of $\Phi_{\rm m} = 0.05$ while the fluid packing fraction $\Phi_{\rm f}$ is taken from the set $\{0.05, 0.1, 0.2, 0.3, 0.5 \}$;
    \item {\bf path II}: the matrix has an intermediate packing fraction of $\Phi_{\rm m} = 0.10$ while the fluid packing $\Phi_{\rm f}$ is taken from the set $\{ 0.05, 0.1, 0.2, 0.3, 0.4, 0.45 \}$;    
    \item {\bf path III}: the matrix has a rather large packing fraction of $\Phi_{\rm m} = 0.20$ while the fluid packing fraction $\Phi_{\rm f}$ is taken from the set  $ \{ 0.05, 0.1, 0.15, 0.2, 0.3 \}$;
    \item {\bf path IV} the fluid has a small packing fraction of $\Phi_{\rm f} = 0.10$ while the matrix packing fraction $\Phi_{\rm m}$ is taken from the set $\{0.05, 0.1, 0.2, 0.25 \}$.
\end{itemize}

Note that along each of these pathways the total packing fraction (i.e. $\Phi_{\rm tot} =\Phi_{\rm f} + \Phi_{\rm f}$) is less than 0.55, a value which is close to the coexistence density between the liquid and the FCC phase of hard spheres. 

In Fig. \ref{fig:pathways} we also show (as a reminder) the ``dynamic'' phase diagram  of spherical particles, confined in a porous matrix, formed again by spherical particles; this information should only be considered as a help of orientation -- for more details cf. Fig. 1 of Ref. \cite{Kurzidim:2009}.

Averages $\langle A \rangle$ were calculated by averaging along simulations with up to $5 \times 10^6$ sweeps (as defined above), while averages $\overline{\langle A \rangle}$ were obtained by averaging over five different (but equivalent) matrix configurations: 
while in the related investigations with spherical particles ten independent matrix configurations were used for the averaging procedure, numerical limitations have forced us to reduce the number of averaging processes by a factor of two: one one side it is the higher numerical costs to identify particle overlap of the spheroids, on the other side it is the fact that we now have to include with $\beta \kappa$ an additional system parameter:

The protocol of the simulations can be summarized as follows. (i) First, we created the matrix configurations: for a given value of $\Phi_{\rm m}$ we performed simulations for $N_{\rm m}$ spherical particles in a simulation box of volume $V$: starting from a random initial configuration, the system was simulated until equilibration was achieved; such an equilibrated particle arrangement was stored as a possible matrix configuration; continuing the simulation over an adequate number of sweeps (guaranteeing that correlations between the different matrix configurations are avoided), equivalent matrix configurations have been created and stored for further investigations. In each of these matrix configurations the particle positions were henceforward kept fixed. (ii) Then the initially spherical fluid particles were inserted into the voids of the matrix configurations. (iii) Finally, the simulations for the full system were launched, applying the above mentioned ``moves'' to the fluid particles, while keeping the positions and the spherical shapes of the matrix particles fixed. Once these QA mixtures were equilibrated we started to record the positions of the fluid particles which eventually led us to the static and dynamic properties of the system.

\begin{figure}[htbp]
\begin{center}
\includegraphics[height=6.5cm]{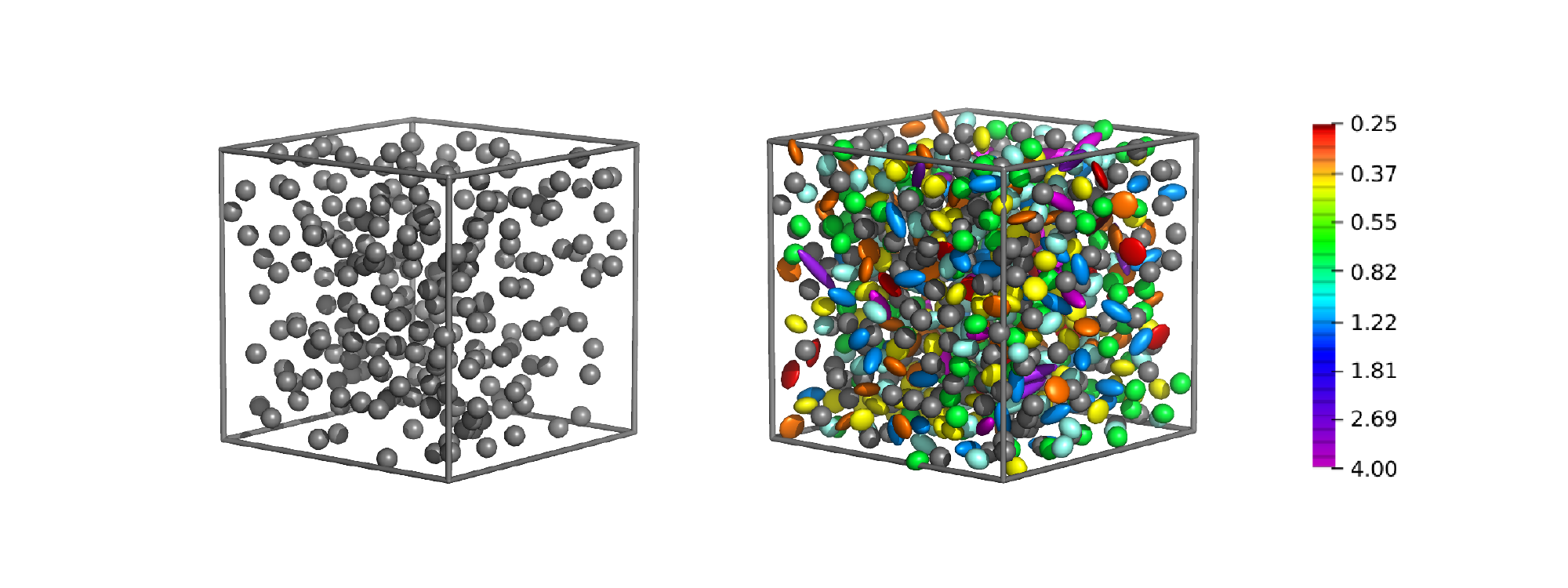}
\caption{(color online) Snapshot of our system with a matrix packing fraction $\Phi_{\rm m} = 0.10$, a fluid packing fraction $\Phi_{\rm f} = 0.2$, and a stiffness value of $\beta \kappa = 10$: the left panel shows the bare configuration of the spherical matrix particles (in grey) while the right panel displays the entire system, i.e., matrix particles in grey and deformed fluid particles in colours. The latter ones are coloured according to the value of their aspect ratio, $x$, using the colour code displayed at the very right: $x = 1$ corresponds to spherical particles, while $x < 1$ and $x > 1$ characterize oblate and prolate ellipsoids of revolution, respectively.}
\label{fig:model}
\end{center}
\end{figure}

\section{Results}
\label{sec:results}

In the following we discuss the results obtained along the four paths through the 
($\Phi_{\rm m}, \Phi_{\rm f}$)-plane, as specified above and as visualized in Fig. \ref{fig:pathways}. Our discussions are based on the radial distribution function, $g(r)$, the probability distribution of the aspect ratio $x$ of the fluid particles, the so-called aspect ratio distribution (termed ARD), the mean squared displacement (MSD), $\delta r^2(t)$, of the fluid particles, and the effective exponent of the MSD, $z(t)$, defined via

\begin{equation}
    z(t) = \frac{d [\log \delta r^2(t)]}{d [\log t]} .
    \label{exponent}
\end{equation}

Throughout the time $t$ is ``measured'' in terms of MC-sweeps.     

\subsection{Path I and path II} 
\label{subsec:path_I_II}

Since results obtained for our system along path I and II display only small quantitative differences we discuss these data together in one subsection. 

\begin{figure}[htbp]
\begin{center}
\includegraphics[width=10cm]{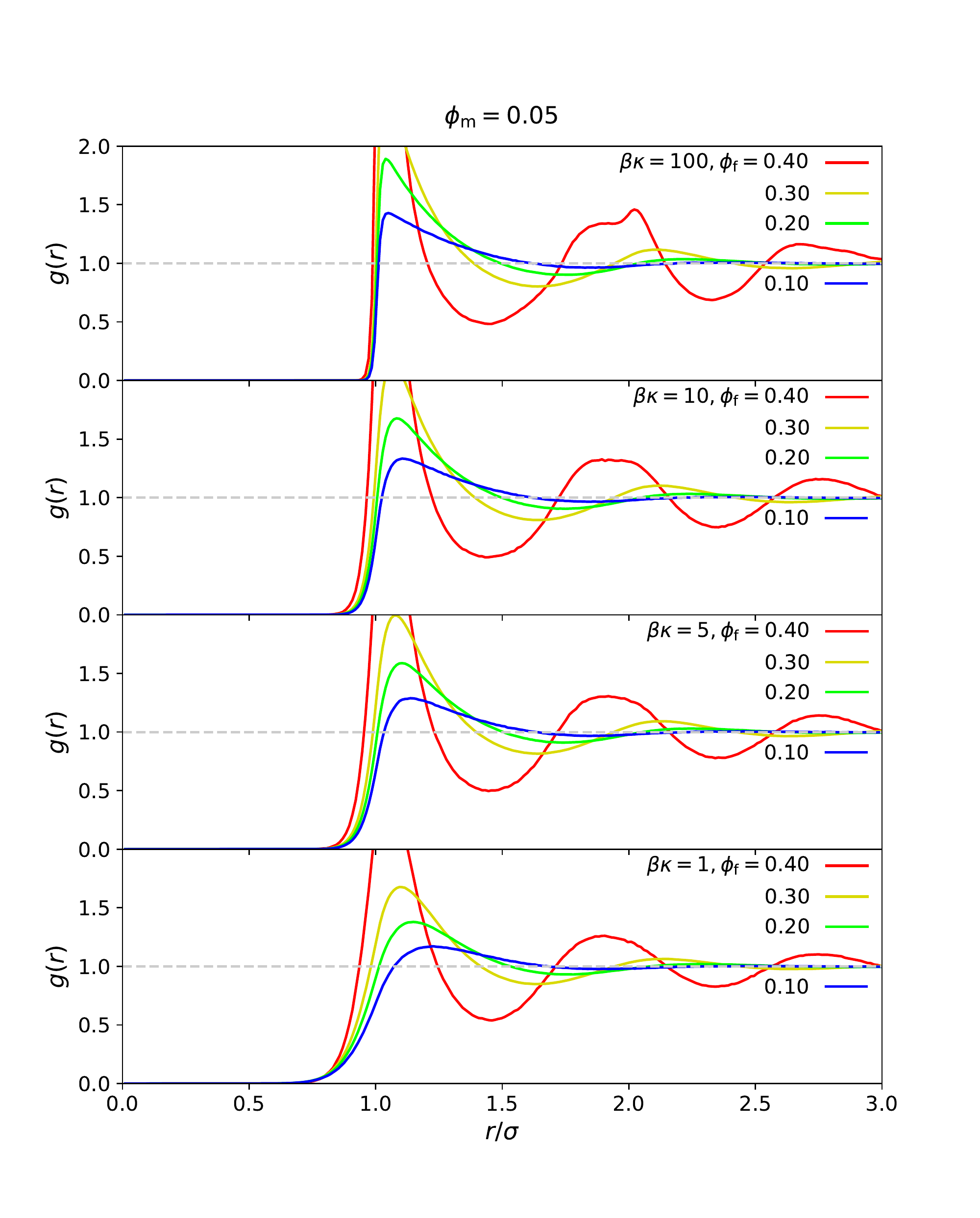}
\caption{(color online) Radial distribution function, $g(r)$, as a function of distance $r$ (in units of $\sigma$) for the fluid particles of our system, calculated along path I and considering various combinations of the fluid packing fraction, $\phi_{\rm f}$, and the stiffness parameter, $\beta \kappa$ (as labeled).}
\label{fig:RDF_I}
\end{center}
\end{figure}

Along these pathways the packing fraction of the matrix assumes rather low values, namely $\Phi_{\rm m} = 0.05$ and $\Phi_{\rm m} = 0.10$, respectively. In Fig. \ref{fig:RDF_I} we have summarized the results for $g(r)$ obtained for states along path I: now that the centers of two particles can approach each other -- depending on their shape -- to distances smaller than $\sigma$ the onset of the main peak of $g(r)$ can assume values that are  smaller than $\sigma$. As a consequence the typical discontinuous peak in $g(r)$ as observed in a hard sphere system, is now a steady function of the distance $r$: for obvious reasons this softening is the more pronounced the smaller the value of $\beta \kappa$, i.e., the stronger the particles are deformable (see also the discussion of the ARD below); in contrast, for $\beta \kappa = 100$ the main peak in the $g(r)$ is still very similar to the one found in a pure hard sphere system. Further, also the onset of this peak shows a substantial shift to smaller distances with decreasing $\beta \kappa$ which is essentially independent of the fluid packing fraction $\Phi_{\rm f}$: for $\beta \kappa = 1$, for instance, the onset of the main peak occurs already at $r \simeq 0.7~ \sigma$. Analysing the peak position in $g(r)$ we note that for $\beta \kappa = 100$ this position is essentially unaffected by the fluid packing fraction, while for strongly deformable particles (e.g., for $\beta \kappa =1$) a clear shift in the position of the peak of $g(r)$ towards larger distances with increasing $\Phi_{\rm f}$ is observed: the strong increase in the preferred distances between strongly deformable particles indicate that they tend to occupy the available space within the matrix more efficiently; those regions are -- for geometric reasons -- not accessible for less deformable (or even rigid, spherical) particles, as they are not able to access narrow pores within the matrix. To conclude we point out that for the highest value of $\Phi_{\rm f}$ and for the highest value of $\beta \kappa$ a shoulder in the side peak of the radial distribution function occurs: this feature indicates the onset of a glassy state, a feature which is agreement with the related investigations of spherical particles confined in a disordered matrix of spherical particles (and widely discussed in Refs.  \cite{Kurzidim:2009,Kurzidim:2011}). Since this shoulder is not observed for more deformable particles (see the other panels of Fig. \ref{fig:RDF_I} for smaller values of $\beta \kappa$) we conclude that a strong deformability of the mobile particles prevents our system from forming a glassy state.

\begin{figure}[htbp]
\begin{center}
\includegraphics[width=10cm]{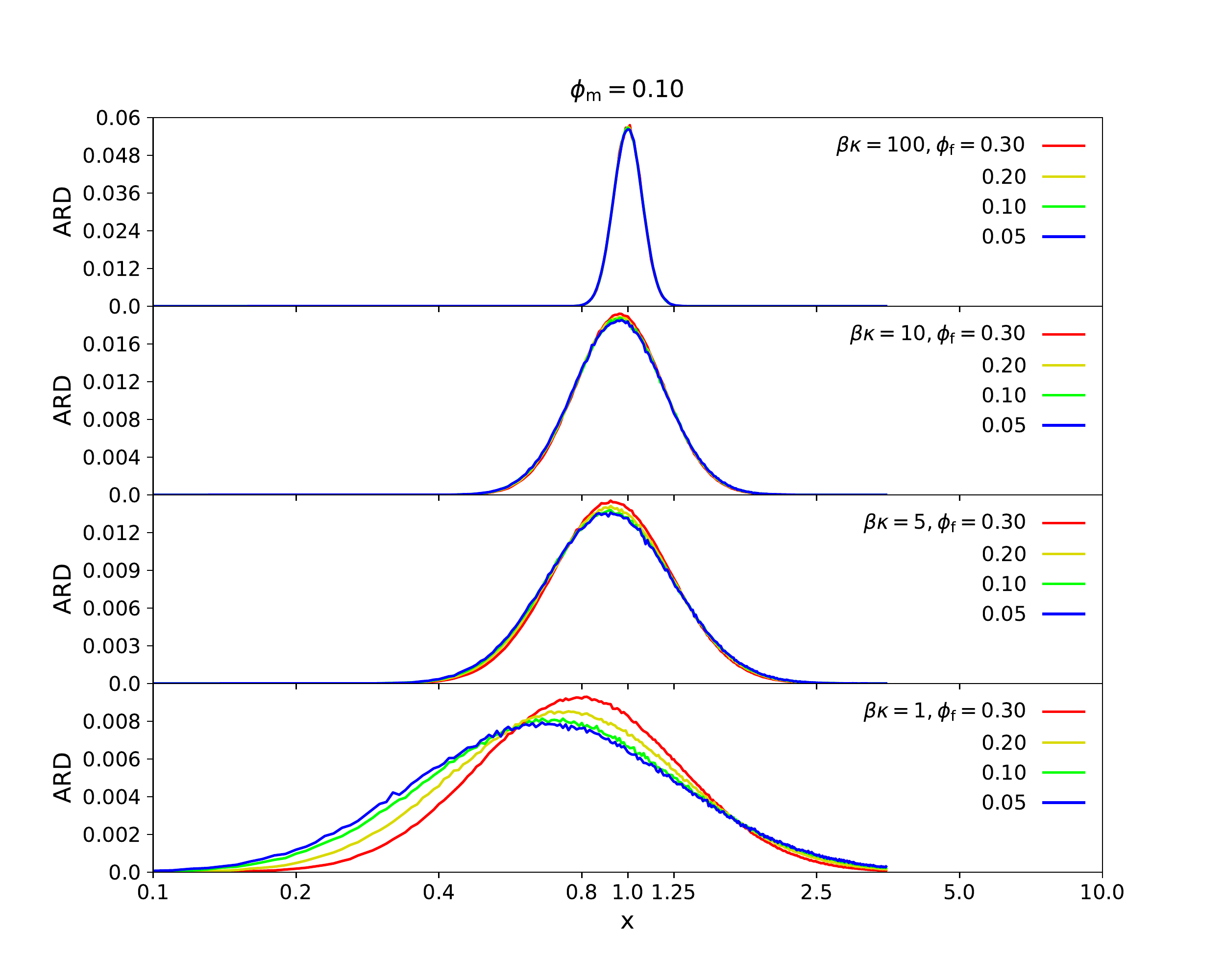}
\caption{(color online) Distribution of the aspect ratio of the fluid particles, ARD($x$), as a function of aspect ratio $x$ for the fluid particles of our system, calculated along path II and considering various combinations of the fluid packing fraction, $\Phi_{\rm f}$, and the stiffness parameter, $\beta \kappa$ (as labeled).}
\label{fig:ARD_II}
\end{center}
\end{figure}

Several of the above features are nicely and consistently complemented by analysing the data of the ARD, which we now discuss along path II; the related data are summarized in Fig. \ref{fig:ARD_II}. These results indicate in an unambiguous manner how the fluid particles do deform under the influence of the surrounding matrix and of other the fluid particles. Our observations can be summarized as follows: the shape and the width of the ARD as a function of the aspect ratio $x$ is primarily triggered by the stiffness parameter $\beta \kappa$, while on a more subordinated level these features are influenced by the available space, notably by $\Phi_{\rm f}$: stiff particles (for instance with $\beta \kappa =100$) show an ARD that is strongly peaked around $x =1$ with a shape that essentially does not depend on $\Phi_{\rm  f}$. As the particles become more deformable (e.g., for $\beta \kappa = 10$ or 5) the ARD is still centered around a value close to $x \simeq 1$, however, the shape of the distribution function broadens considerably; still no sizable shape dependence on $\Phi_{\rm f}$ can be observed. Only for strongly deformable particles (i.e., for $\beta \kappa =1$) the impact of the fluid packing fraction $\Phi_{\rm f}$ becomes visible: (i) we find very broad ARDs whose peak positions have shifted to smaller $x$-values -- e.g. for $\Phi_{\rm f} = 0.05$ we find $x_{\rm max} \simeq 0.6$ (corresponding to pronounced oblate particles); (ii) concomitantly, the width of the ARD broadens strongly with decreasing fluid packing fraction: for $\beta \kappa = 1$ and $\Phi_{\rm f} = 0.05$ the ``wings'' of the ARD still assumes sizable values for $x \simeq 0.2$ and $x \simeq 2.5$: for obvious reasons particles with such extremely small aspect ratios are now able to populate narrow pores inside the matrix which are inaccessible to particles if they were spherical  -- which brings us consistently back to our above discussion of $g(r)$.


\subsection{Path III} 
\label{subsec:path_III}

For states located along path III a relatively high matrix packing fraction of $\Phi_{\rm m} = 0.20$ is assumed, while the fluid packing fraction reaches values of up to $\Phi_{\rm f} = 0.20$. 

The radial distribution functions show for states located along path III similar features -- i.e., the onset, the softening and the shift of the main peak in $g(r)$ -- as the data discussed in Subsection \ref{subsec:path_I_II}; therefore we refrain from showing these results but rather refer the reader to  Figs. \ref{fig:RDF_all_100} and \ref{fig:RDF_all_5} in Subsection \ref{subsec:across_pathways} and the related discussion. Furthermore we report that the ARDs of systems along path III show a qualitatively similar behaviour (with minor qualitative differences) as the data discussed along pathways I and II; thus we avoid an extensive discussion of the ARDs along path III.

\begin{figure}[htbp]
\begin{center}
\includegraphics[width=10cm]{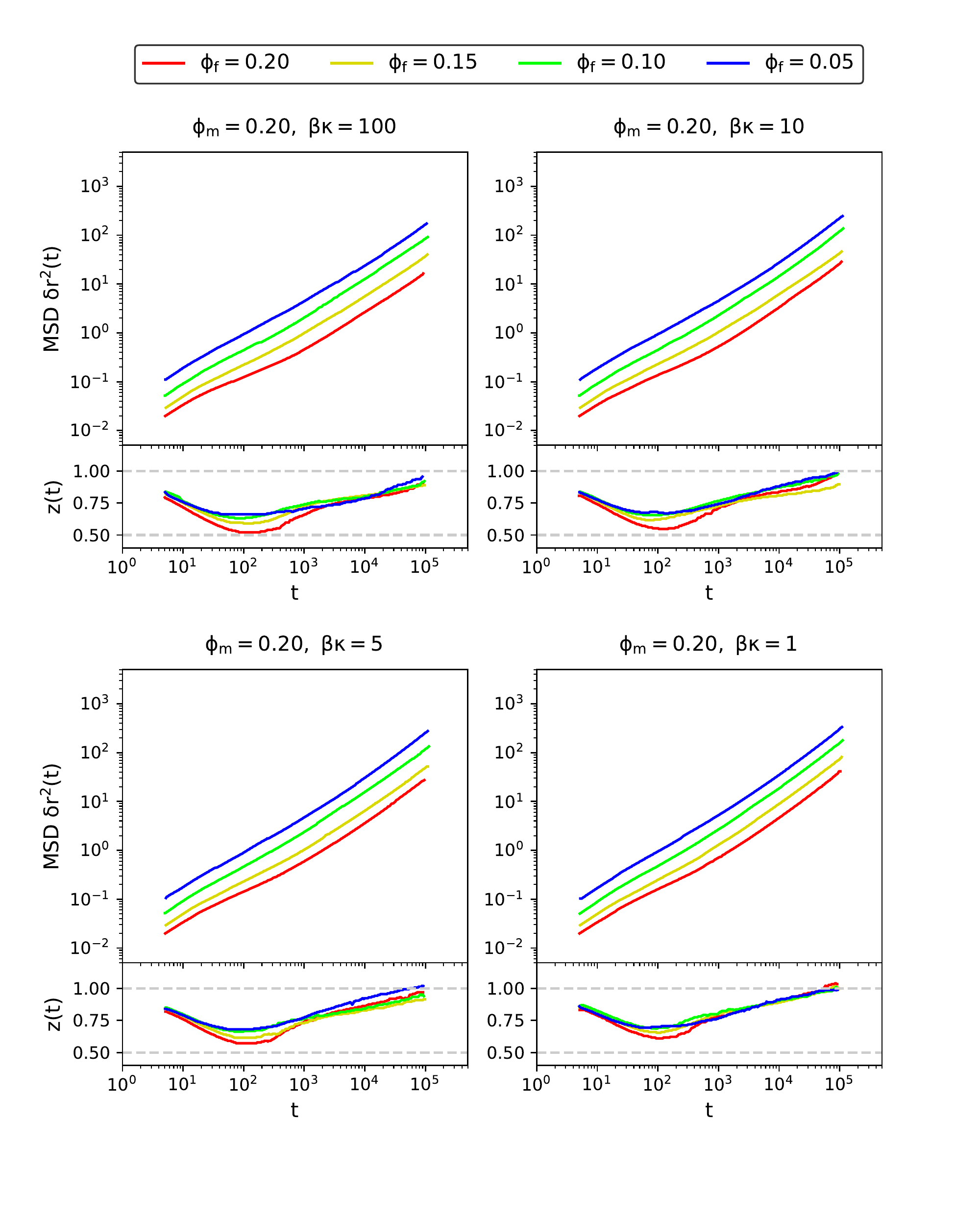}
\caption{(color online) Mean squared displacement (MSD), $\delta r^2(t)$, as a function of ``time'' $t$ (in terms of MC sweeps) for the fluid particles of our system, calculated along path III and considering various combinations of the fluid packing fraction, $\Phi_{\rm f}$, and the stiffness parameter, $\beta \kappa$ (as labeled).} 
\label{fig:MSD_III}
\end{center}
\end{figure}

Instead we focus in the following on a detailed analysis  of the MSD, $\delta r^2(t)$, with the time $t$ measured in MC sweeps and shown in Fig. \ref{fig:MSD_III}; results are collected in panels according to the $\beta \kappa$-values. For the long-time diffusivity (i.e., for $10^4 \lesssim t$) we can make the following observations: the MSDs reach over a comparable time range for a given value of $\Phi_{\rm f}$ -- and essentially irrespective of the $\beta \kappa$-value -- the same value: thus we conclude that under the conditions imposed along path III the diffusivity of the particles depends solely on $\Phi_{\rm f}$. In contrast, for intermediate time ranges (i.e., for $10^1 \lesssim t \lesssim 5~10^3$) we observe a pronounced subdiffusive behaviour: it is characterized by values of the effective exponent of the MSD, $z(t)$, as defined in Eq. (\ref{exponent}), that are considerably smaller than unity. In  such a scenario particles are temporarily trapped in local cages, formed either by the surrounding fluid particles or by voids of the matrix. The degree of subiffusivity is primarily the stronger the less deformable the particles are: notably we find for $\beta \kappa = 100$ values of the exponent as small as $z \simeq 0.5$. On the other hand, particles with small $\beta \kappa$-values can obviously escape -- thanks to their strong deformability -- those cages much easier: consequently $z(t)$ recovers a value of unity at considerably shorter times (as compared to systems with large $\beta \kappa$-values). On a secondary level, the degree of subdiffusivity is triggered by the fluid density $\Phi_{\rm f}$: for a  given $\beta \kappa$-value this dependence is the more pronounced the denser the fluid particles. 




\subsection{Path IV} 
\label{subsec:path_IV}

Along path IV we now keep the fluid density $\Phi_{\rm f}$ fixed and vary the matrix density $\Phi_{\rm m}$ over a rather broad range: thus, with increasing $\Phi_{\rm m}$ we steadily increase the confinement of the mobile particles.

\begin{figure}[htbp]
\begin{center}
\includegraphics[width=10cm]{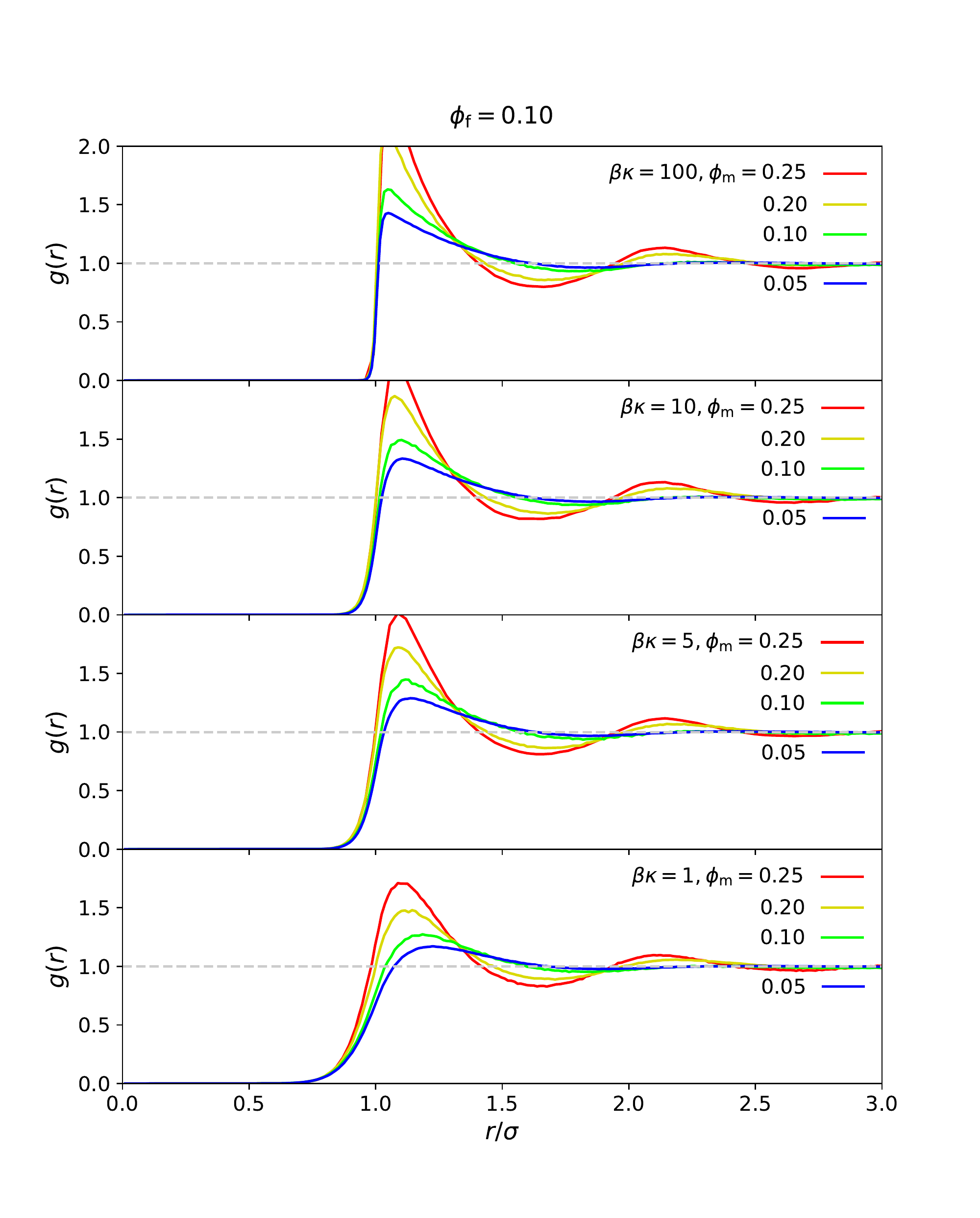}
\caption{(color online) Radial distribution function, $g(r)$, as a function of distance $r$ (in units of $\sigma$) for the fluid particles of our system, calculated along path IV and considering various combinations of the fluid packing fraction, $\phi_{\rm f}$, and the stiffness parameter, $\beta \kappa$ (as labeled).}
\label{fig:RDF_IV}
\end{center}
\end{figure}

\begin{figure}[htbp]
\begin{center}
\includegraphics[width=10cm]{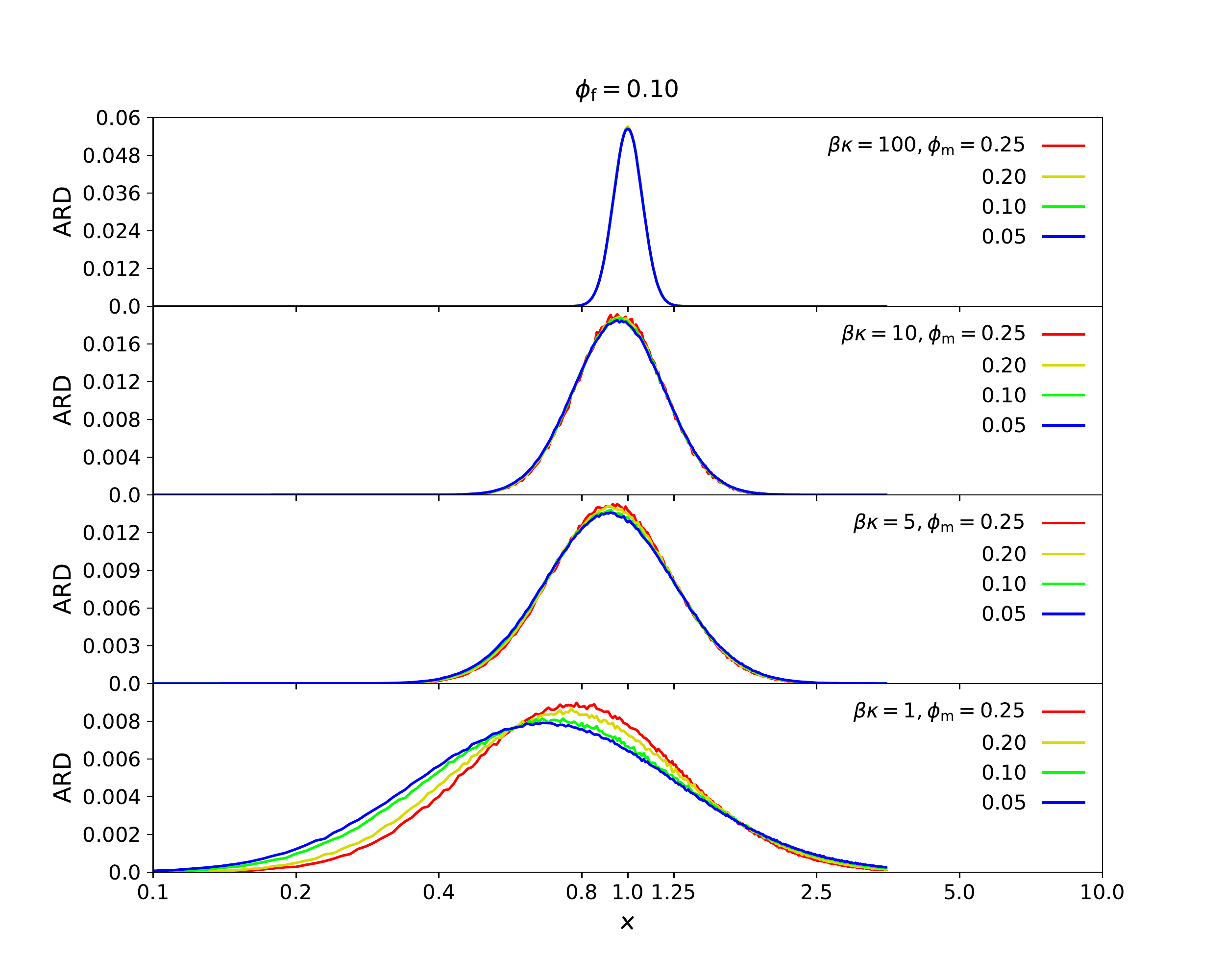}
\caption{(color online) Distribution of the aspect ratio of the particles, ARD($x$), as a function of aspect ratio $x$ for the fluid particles of our system, calculated along path IV and considering various combinations of the fluid packing fraction, $\Phi_{\rm f}$, and the stiffness parameter, $\beta \kappa$ (as labeled).}
\label{fig:ARD_IV}
\end{center}
\end{figure}

Data for the radial distribution function are shown in Fig. \ref{fig:RDF_IV}. Similar as for the related data collected along the other pathways we observe that the deformability of the particles induces with decreasing $\beta \kappa$ a strong softening of the main peak in $g(r)$ and a considerable shift of the onset of this peak towards smaller distances. The related data for the ARD are shown in Fig. \ref{fig:ARD_IV}. Compared to the preceding results we find similar features and dependencies: (i) the lower the value of $\beta \kappa$, the broader the ARD as a function of the aspect ratio $x$; (ii) only for strongly deformable particles (i.e., for $\beta \kappa = 5$ or even smaller) the packing of the particles (now imposed by the packing fraction of the matrix, $\Phi_{\rm m}$) starts to have a distinct influence on the shape of the ARD and its peak position shifts with decreasing $\Phi_{\rm m}$ to smaller $x$-values, located at $x \simeq 0.55$. 

\begin{figure}[htbp]
\begin{center}
\includegraphics[width=10cm]{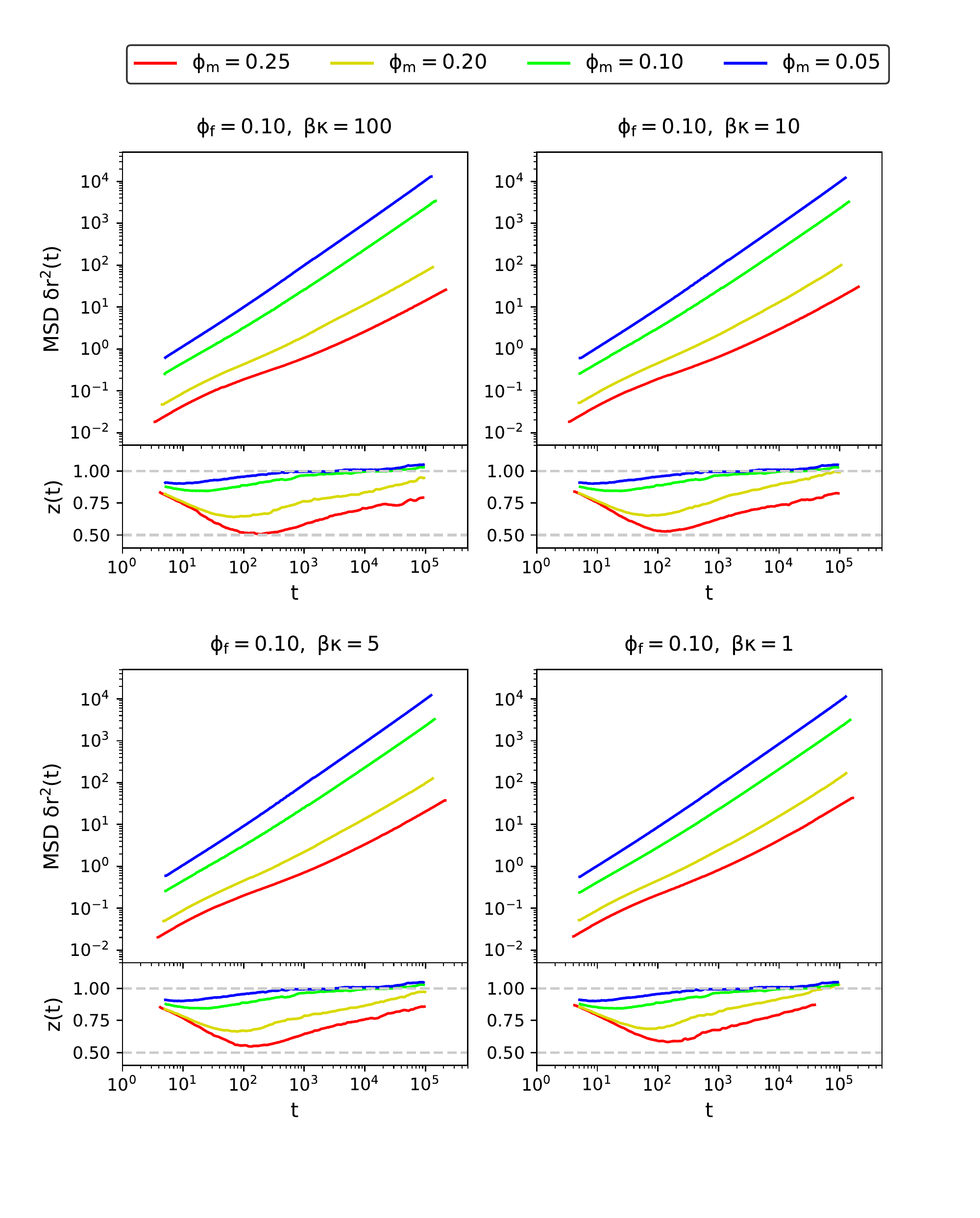}
\caption{(color online) Mean squared displacement (MSD), $\delta r^2(t)$, as a function of ``time'' $t$ (in terms of MC sweeps) for the fluid particles of our system, calculated along path IV and considering various combinations of the fluid packing fraction, $\Phi_{\rm f}$, and the stiffness parameter, $\beta \kappa$ (as labeled).}
\label{fig:MSD_IV}
\end{center}
\end{figure}

For the MSDs (shown in Fig. \ref{fig:MSD_IV}) we identify distinctive differences as compared to the data accumulated for states along path III (see Fig. \ref{fig:MSD_III}). We start again with the long-time diffusivity (typically encountered for $10^4 \lesssim t$). Again the main impact on the MSD originates from the packing fraction of the matrix: as we increase $\Phi_{\rm m}$ from 0.05 to 0.20 (i.e., by a factor of four), the distance that the particles are able to cover over a comparable time range differs by more than three orders of magnitude (compare differently coloured curves in each of the MSD-panels of Fig. \ref{fig:MSD_IV}). In contrast, the MSD  essentially does not depend on the deformability: particles cover for a given $\Phi_{\rm f}$-value and again after a comparable time span essentially the same distance, irrespective of the deformability $\beta \kappa$ (compare equally coloured curves in the different MSD-panels of Fig. \ref{fig:MSD_IV}). Thus we conclude that the long-term diffusivity of the particles within the surrounding matrix is triggered by confinement, while the deformability plays (at least for the value of $\Phi_{\rm f}$ that we have chosen) only a negligible role. From the slope of the MSD in the linear regime we can read off the diffusion constant: we observe that for those system parameters where the diffusive behaviour has been regained (i.e., where the exponent $z(t)$ has recovered values close to unity) the diffusion constants are different for the different values of $\Phi_{\rm m}$, in striking contrast to path III. Eventually, at intermediate time scales (i.e., $10^1 \lesssim t \lesssim 5~10^3$) the packing fraction of the matrix, $\Phi_{\rm m}$ is again the decisive factor for the subdiffusive behaviour; still (and in contrast to path III) the deformability has a more relevant impact: depending on the value of $\beta \kappa$ we observe pronounced differences in the exponent $z(t)$ (compare equally coloured curves in the different $z(t)$-panels of Fig. \ref{fig:MSD_IV}): for strongly deformable particles (e.g., $\beta \kappa = 1$) the exponent $z(t)$ is -- in particular for the higher matrix concentrations -- typially by 10 \% smaller than in the case where the fluid particles are essentially undeformable (i.e., $\beta \kappa = 100$). Also the subdiffusive regime extends -- as compared to path III -- over a considerably broader time range: notably at the highest matrix concentration, i.e., for $\Phi_{\rm m} = 0.25$, the exponent $z(t)$ assumes for the observed time window a value which ranges considerably below unity.



\subsection{The radial distribution functions across the pathways}
\label{subsec:across_pathways}

\begin{figure}[htbp]
\begin{center}
\includegraphics[width=10cm]{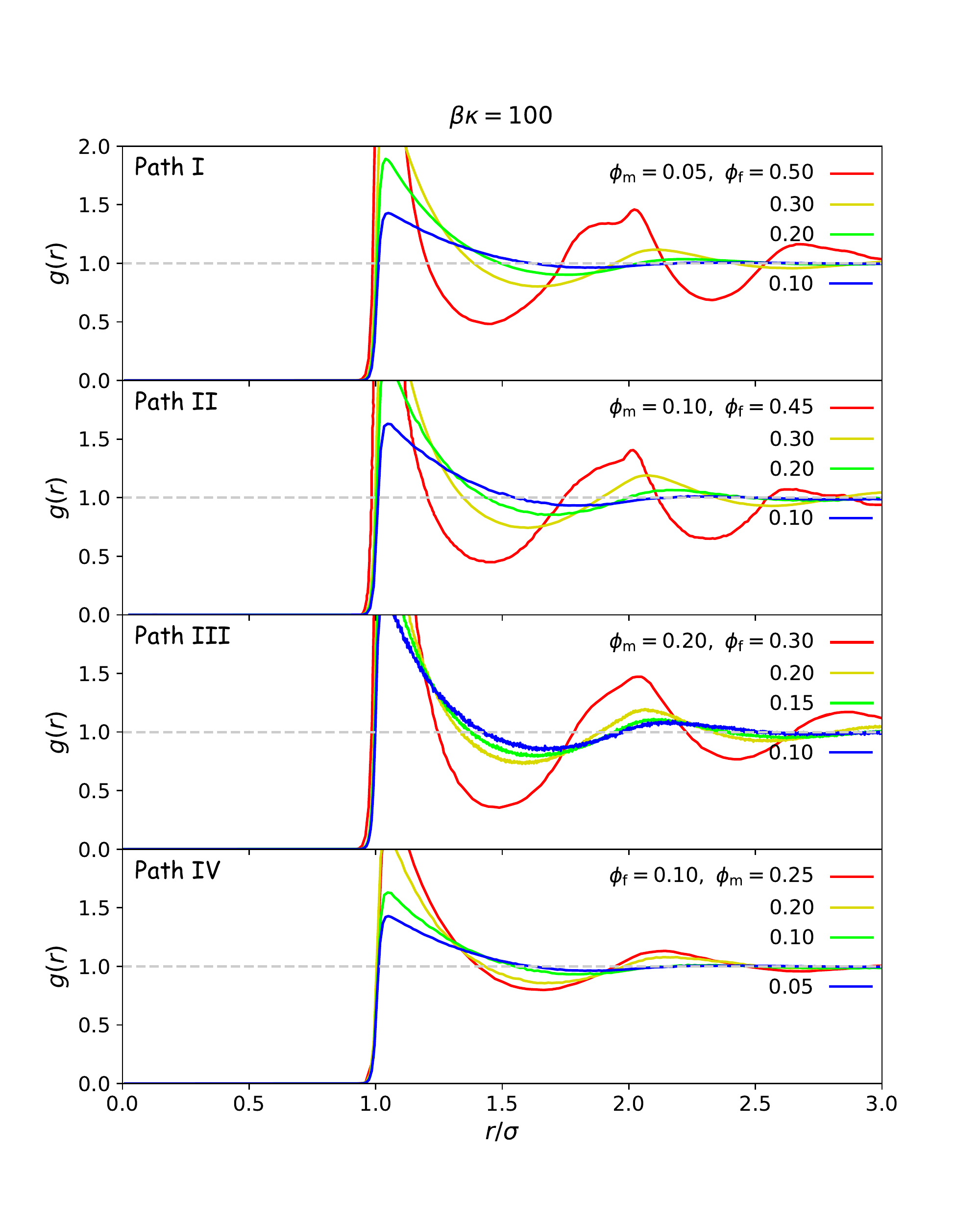}
\caption{(color online) Radial distribution function, $g(r)$, as a function of distance $r$ (in units of $\sigma$) for the fluid particles of our system, calculated along each of the four pathways (as labeled) and considering various combinations of the fluid and the matrix packing fractions, $\Phi_{\rm f}$ and $\Phi_{\rm m}$, respectively (as labeled). The stiffness parameter assumes for all systems the value $\beta \kappa = 100$.}
\label{fig:RDF_all_100}
\end{center}
\end{figure}

\begin{figure}[htbp]
\begin{center}
\includegraphics[width=10cm]{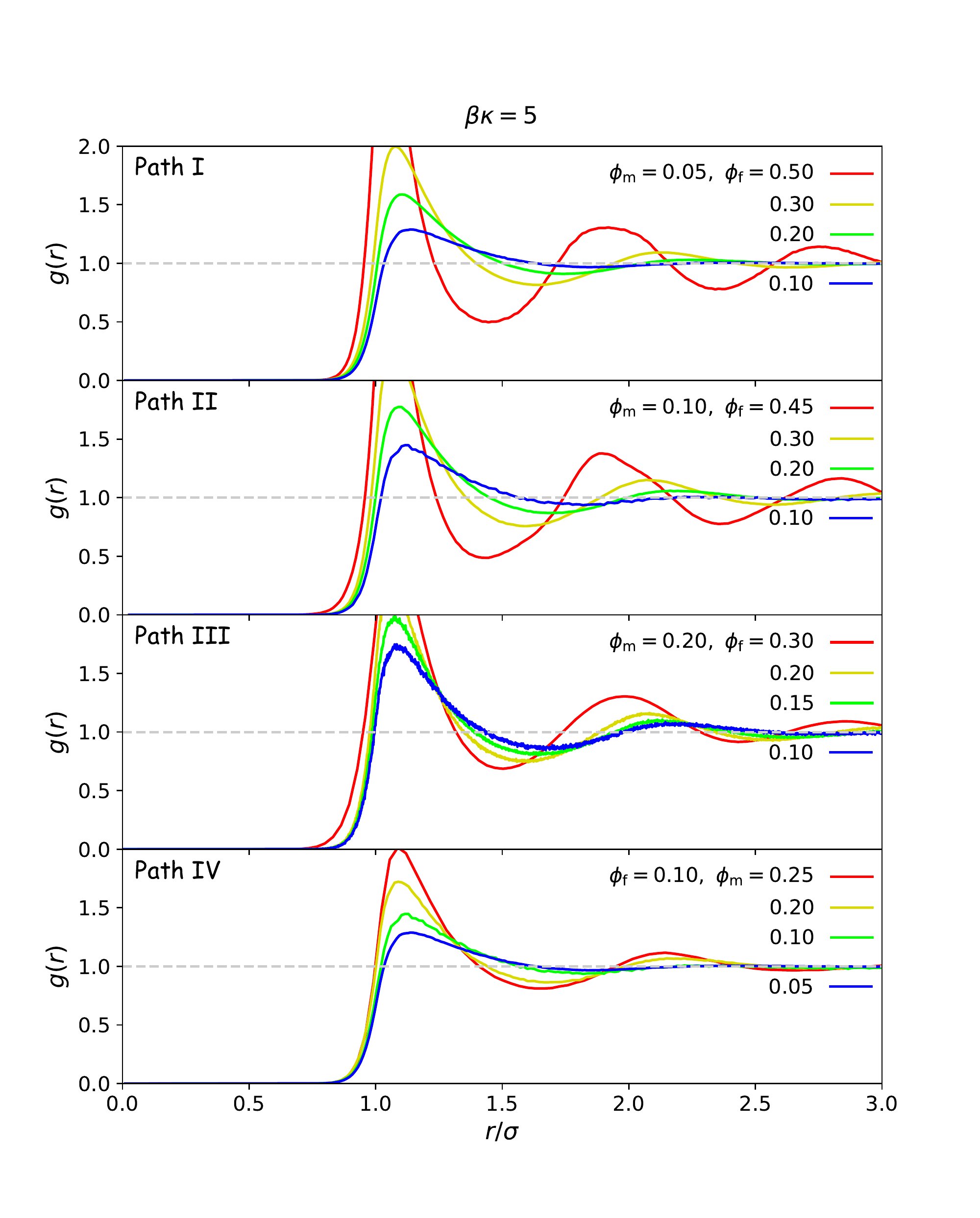}
\caption{(color online) Radial distribution function, $g(r)$, as a function of distance $r$ (in units of $\sigma$) for the fluid particles of our system, calculated along each of the four pathways (as labeled) and considering various combinations of the fluid and the matrix packing fractions, $\Phi_{\rm f}$ and $\Phi_{\rm m}$, respectively (as labeled). The stiffness parameter assumes for all systems the value $\beta \kappa = 5$.}
\label{fig:RDF_all_5}
\end{center}
\end{figure}

To conclude our analysis we have collected the radial distribution functions along {\it all} pathways for hardly (i.e., $\beta \kappa =100$) and for strongly (i.e., $\beta \kappa = 5$) deformable particles and for all density parameters available in Figs. \ref{fig:RDF_all_100} and \ref{fig:RDF_all_5}, respectively. The aim of this discussion is to disentangle the impact of the deformability, $\beta \kappa$, from the density parameters, $\Phi_{\rm m}$ and $\Phi_{\rm f}$, on the shape of the distribution function. 

From the accumulated data it is obvious that $\beta \kappa$ is primarily responsible for the shape of $g(r)$, in particular for the main peak: a large value of $\beta \kappa$  (Fig. \ref{fig:RDF_all_100}) leads to a harshly repulsive peak with an onset close to  $\sigma$; the position of this peak changes only marginally as either of the densities (i.e., $\Phi_{\rm m}$ or $\Phi_{\rm f}$) are changed. Side shoulders in the second peak of $g(r)$ for systems with high total packing fraction ($\Phi_{\rm tot} = \Phi_{\rm f} + \Phi_{\rm m}$) indicate the occurrence of a glassy state for the case of hardly deformable fluid particles.  In striking contrast, we find for strongly deformable particles (i.e., $\beta \kappa = 5$, with data shown in Fig. \ref{fig:RDF_all_5}) that the main peak of $g(r)$ is now a smooth function with an onset at distances $r \simeq 0.8 \sigma$ (or even smaller -- cf. Figs. \ref{fig:RDF_I} or \ref{fig:RDF_IV}). But also the position of the main peak is no longer restricted to values close to $\sigma$, but can reach -- depending on the density parameters -- values as high as $r \simeq 1.25 \sigma$. We conclude that the strong deformability of the particles which can either lead to center-to-center distances between particles that are smaller than $\sigma$, but also to separations that are larger than $\sigma$. Possibly the particles can also explore the available space inside the pores of the matrix more efficiently as they can access in their deformed shape spaces that otherwise would be inaccessible for rigid, spherical particles; however, a closer analysis of this conjecture would require a detailed geometric analysis of the voids inside the matrix in terms of a Delaunay decomposition \cite{Kurzidim:2011a}, an investigation which would clearly bypass the limitations of this contribution. 

Of course also the density parameters, $\Phi_{\rm f}$ and $\Phi_{\rm m}$, have their impact on the shape of the radial distribution function. Interestingly, if one compares the $g(r)$ of systems with (approximately) the same total packing fraction, $\Phi_{\rm tot}$, and given $\beta \kappa$ the respective curves differ only marginally in their shape; thus we conclude that the shape of $g(r)$ primarily depends on the total packing of the system and that the parition of this quantity  into matrix and fluid densities plays only a minor role. Possibly similar conclusions were also available in the preceding study of spherical fluid particles in a matrix of hard spheres \cite{Kurzidim:2009,Kurzidim:2010,Kurzidim:2011}, but have not been investigated more closely.

\section{Conclusions}
\label{sec:conclusions}

We have investigated in extensive Monte Carlo simulations the properties of impenetrable, deformable, fluid particles that are immersed into a matrix formed by immobile, impenetrable spherical particles. Using the concept of deformable particles, put forward by Batista and Miller \cite{Batista:2010,Batista:2011}, the mobile constituents of the system can modify their shape within the class of ellipsoids of revolution. The energy change related to this deformation is fed into suitably adapted Monte Carlo selection rules; thus we apply in our simulations -- apart from the translational and orientational moves -- also shape deformation moves for the fluid particles. Overlap between the (fluid and matrix) particles was detected by using a criterion derived by Vieillard-Baron \cite{Vieillard-Baron:1972} for elliptic particles; we note that from the computational point of view this criterion is computationally considerably more expensive than the one for simple spherical particles. The above mentioned energy change related to the shape deformation is a function of the aspect ratio $x$ of the particles; its strength is measured in terms of the deformability $\beta \kappa$: for $\beta \kappa \to \infty$ we recover undeformable hard spheres, while small $\beta \kappa$ values specify easily deformable particles. 

Our investigations are based on the quenched-annealed concept \cite{Madden:1988,Madden:1992,Given:1992,Given:1992a,Given:1994} where physical quantities are evaluated from the particle positions and orientations via a double averaging procedure: (i) in a first step a trace is taken over the degrees of freedom of the fluid particles for a given matrix configuration; (ii) in a second step observables are averaged over a limited number of different, but equivalent matrix configurations; the high numerical cost imposed by the overlap criterion of elliptic particles forced us to limit the second averaging process to five independent matrix configurations per state point investigated. In total, the system is characterized by the packing fractions of the mobile (fluid) and the immobile matrix particles ($\Phi_{\rm f}$ and $\Phi_{\rm m}$, respectively) and the deformability parameter $\beta \kappa$.

In an effort to extract as much information as possible we have defined in the parameter space spanned by $\Phi_{\rm f}$ and $\Phi_{\rm m}$ four pathways which were also used in a previous investigation of spherical fluid particles confined in a matrix of spherical particles \cite{Kurzidim:2009,Kurzidim:2010,Kurzidim:2011}. The pathways are characterized by keeping either of the aforementioned densities constant, while varying the other density over a representative range. For the specification of these ranges we dwelled on the previous investigation which provided some insight into the density range where the system remains in a disordered -- i.e., either liquid or at most glassy -- state. Our investigations and our analysis are based on the radial distribution function, $g(r)$, the distribution of the aspect ratio (ARD), the spatial mean squared displacement (MSD), $\delta r^2(t)$, and its effective exponent, $z(t)$. 

The shape of the radial distribution function is primarily influenced by the deformability of the particles: in systems of strongly deformable, mobile particles the main peak in $g(r)$ is a smooth function of the distance and its onset can be located at distances as 0.7 $\sigma$. In contrast, hardly deformable particles have harsh, hard-particle type main peaks which are located essentially at $\sigma$. For a given value of $\beta \kappa$ the shape of $g(r)$ is imposed by the total packing of the system.  The ARD is a strongly peaked function for high values of deformability and extends over a growing $x$-range as  $\beta \kappa$ decreases; only for strongly deformable particles an impact of the packing of the system is observable: for very low densities the ARD can extend over a remarkably broad range of $x$-values.

%
%
%
%
%
%

\section*{Acknowledgments}  The authors acknowledge financial support by the Austrian Science Foundation (FWF) under projects No. I3846-N36 and by E-CAM, an e-infrastructure center of excellence for software, training and consultancy in simulation and modelling funded by the EU (Proj. No. 676531). Helpful and enlightening discussions with Mark Miller (Durham) are gratefully acknowledged. We thank Patrick Chalupa (Vienna) for technical help.








\end{document}